\documentclass[fleqn]{article}
\usepackage{amsmath,amssymb}
\usepackage{epsfig,graphicx}

\catcode`@=11 \@addtoreset{equation}{section} \catcode`@=12

\begin{document}
\title{\vskip-1.7cm \bf  CFT driven cosmology and conformal higher spin fields}
\date{}
\author{A.O.Barvinsky}
\maketitle
\hspace{-5.5mm}\centerline{\em Theory Department, Lebedev
Physics Institute, Leninsky Prospect 53, Moscow 119991, Russia}\\
\centerline{\em Department of Physics, Tomsk State University, Lenin Ave. 36, Tomsk 634050, Russia}\\
\centerline{\em and}\\
\centerline{\em Pacific Institue for Theoretical Physics, Department of Physics and Astronomy, University }\\
\centerline{\em of British Columbia, 6224 Agricultural Road, Vancouver, BC V6T 1Z1, Canada}

\begin{abstract}
Conformal higher spin (CHS) field theory, which is a solid part of recent advanced checks of AdS/CFT correspondence, finds applications in cosmology. Hidden sector of weakly interacting CHS fields suggests a resolution of the hierarchy problem in the model of initial conditions for inflationary cosmology driven by a conformal field theory. These initial conditions are set by thermal garland type cosmological instantons in the sub-planckian energy range for the model of CHS fields with a large positive coefficient $\beta$ of the Gauss-Bonnet term in their total conformal anomaly and a large number of their polarizations $\mathbb{N}$. The upper bound of this range $M_P/\sqrt\beta$ is shown to be much lower than the gravitational cutoff $M_P/\sqrt\mathbb{N}$ which is defined by the requirement of smallness of the perturbatively nonrenormalizable graviton loop contributions. In this way we justify the approximation scheme in which the nonrenormalizable graviton sector is subject to effective field theory under this cutoff, whereas the renormalizable sector of multiple CHS fields is treated beyond perturbation theory and dynamically generates the bound on the inflation scale of the CFT cosmology $M_P/\sqrt\beta\ll M_P/\sqrt\mathbb{N}$. This confirms recent predictions for the origin of the Starobinsky $R^2$ and Higgs inflation models from the CHS cosmology, which occurs at the energy scale three or four orders of magnitude below the gravitational cutoff, $\sqrt{\mathbb{N}/\beta}\sim 10^{-3}\div 10^{-4}$. We also consider cosmological models dominated by fermionic CHS fields with a negative $\beta$ and anomaly free models of infinite towers of CHS fields with $\beta=0$ and briefly discuss the status of unitarity in CHS models.
\end{abstract}

\section{Introduction}
Rapidly developing trend in nonperturbative approach to quantum gravity and string theory, which is based on holographic ideas of AdS/CFT correspondence \cite{AdS/CFT0}, involves Vasiliev theory of interacting higher spin fields \cite{Vasiliev} and naturally leads to the notion of conformal higher spin fields (CHS) \cite{GKPST}. Though these CHS fields represent thus far only a playground for rather sophisticated verification of the AdS/CFT correspondence \cite{GKPST,Tseytlin}, quite interestingly they turn out to be important in recent cosmological applications associated with the problem of initial conditions in the early inflationary Universe \cite{slih,why,hill-top}. This is the model of the CFT driven cosmology \cite{slih,why} which incorporates two main ideas -- a new concept of the cosmological microcanonical density matrix as the initial state of the Universe and the implementation of this concept in cosmology with a large number of quantum fields conformally coupled to gravity.

This model plays important role within quantum cosmology and within the cosmological constant and dark energy problems. In particular, it resolves the issue of infrared catastrophe associated with the observer independent treatment of the no-boundary state \cite{noboundary} -- an anti-intuitive conclusion that the origin of an infinitely big universe (with an insufficient amount of inflation produced at the zero minimum of the inflaton potential rather than its maximum) is infinitely more probable than that of a finite one. Another property is that its statistical ensemble is bounded to a finite range of values of the effective cosmological constant \cite{slih} and gives rise to a new type of hill-top inflation \cite{hill-top,slih_inflation}. Also this model incorporates a certain version of the dS/CFT holographic duality \cite{DGP/CFT} and is potentially capable of generating the cosmological acceleration phenomenon within the so-called Big Boost scenario \cite{bigboost}.

The setting of the initial conditions problem is based on canonical quantization of gravity theory and a natural notion of the microcanonical density matrix as a projector on the space of solutions of the quantum gravitational Dirac constraints -- the system of the Wheeler-DeWitt equations \cite{why,whyBFV}. Its statistical sum has a representation of the Euclidean quantum gravity (EQG) path integral \cite{slih,why}
    \begin{eqnarray}
    &&Z=
    \!\!\int\limits_{\,\,\rm periodic}
    \!\!\!\! D[\,g_{\mu\nu},\varPhi\,]\;
    e^{-S[\,g_{\mu\nu},\varPhi\,]},         \label{Z}
    \end{eqnarray}
over metric $g_{\mu\nu}$ and matter fields $\varPhi$ which are
periodic on the Euclidean spacetime with a time compactified to a circle $S^1$. As shown in \cite{slih,why}, this statistical sum has a good predictive power in the model with a primordial cosmological constant $\varLambda$ and the matter sector of a large number $\mathbb{N}$ of quantum fields $\varPhi$ conformally coupled to gravity -- conformal field theory (CFT) with the action $S_{CFT}[\,g_{\mu\nu},\varPhi\,]$,
    \begin{eqnarray}
    &&S[\,g_{\mu\nu},\varPhi\,]=-\frac{M_P^2}2
    \int d^4x\,g^{1/2}\,(R-2\varLambda)
    +S_{CFT}[\,g_{\mu\nu},\varPhi\,].    \label{tree0}
    \end{eqnarray}
What allows one to go beyond a usual semiclassical expansion is that the CFT modes $\varPhi$ dominate over conformally non-invariant fields and spatially-inhomogeneous metric modes. Integration over $\varPhi$ in (\ref{Z}) then leads to the effective action
    \begin{eqnarray}
    &&S_{\rm eff}[\,g_{\mu\nu}]=-\frac{M_P^2}2
    \int d^4x\,g^{1/2}\,(R-2\varLambda)
    +\varGamma_{CFT}[\,g_{\mu\nu}],          \label{eff}\\
    &&e^{-\varGamma_{CFT}[\,g_{\mu\nu}]}=\int D\varPhi\,e^{-S_{CFT}[\,g_{\mu\nu},\varPhi\,]}.
    \end{eqnarray}
It differs from (\ref{tree0}) by $S_{CFT}[\,g_{\mu\nu},\varPhi\,]$ replaced with $\varGamma_{CFT}[\,g_{\mu\nu}]$ -- the effective action of the conformal fields on the background of $g_{\mu\nu}$.

On Friedmann-Robertson-Walker (FRW) background this action is exactly calculable by using the local conformal transformation to the static Einstein universe and the well-known gravitational trace anomaly
    \begin{eqnarray}
    &&g_{\mu\nu}\frac{\delta
    \varGamma_{CFT}}{\delta g_{\mu\nu}} =
    \frac{1}{4(4\pi)^2}g^{1/2}
    \left(\alpha \Box R +\beta E +
    \gamma C_{\mu\nu\alpha\beta}^2\right),               \label{anomaly}
    \end{eqnarray}
where the coefficients of local curvature invariants -- $\Box R$, Gauss-Bonnet term $E=R_{\mu\nu\alpha\gamma}^2-4R_{\mu\nu}^2
+ R^2$ and Weyl tensor squared $C_{\mu\nu\alpha\beta}^2$ -- are determined by the CFT particle content. These coefficients are additive sums of contributions of all conformal fields of different spins $s$. In particular, the coefficient $\beta$ of the Gauss-Bonnet term, which is of major interest in what follows, reads
    \begin{eqnarray}
    \beta=\sum_s\beta_s \,\mathbb{N}_s,                \label{101}
    \end{eqnarray}
where $\beta_s$ is a partial contribution of spin $s$ conformal field and $\mathbb{N}_s$ is the number of such fields in the set of all $\varPhi$.

The resulting $\varGamma_{CFT}[\,g_{\mu\nu}]$ becomes the sum of the anomaly contribution and a free energy of conformal matter fields on the sphere $S^3$ at the temperature determined by the period of the Euclidean time. Then
the physics of the CFT driven cosmology is entirely determined by the effective action (\ref{eff}). Solutions of its equations of motion, which give a saddle point of the statistical sum path integral, are the cosmological instantons of $S^1\times S^3$ topology with the Friedmann-Robertson-Walker metric
    \begin{eqnarray}
    g_{\mu\nu}^{FRW}dx^\mu dx^\nu=N^2(\tau)\,d\tau^2
    +a^2(\tau)\,d^2\Omega^{(3)},              \label{FRW}
    \end{eqnarray}
where a periodic lapse function $N(\tau)$ and scale factor $a(\tau)$ are the functions of the Euclidean time belonging on $S^1$ \cite{slih}. These instantons serve as initial conditions for the cosmological evolution $a_L(t)$ in the physical Lorentzian signature spacetime, which follows from $a(\tau)$ by analytic continuation $a_L(t)=a(\tau_++it)$ at the point of the maximum value of the Euclidean scale factor $a_+=a(\tau_+)$. The coefficient $\beta$ of the Gauss-Bonnet term in (\ref{anomaly}) plays especially important role because it imposes an upper bound on the range of $\varLambda$ within which these instantons exist,
    \begin{eqnarray}
    \varLambda\leq\frac{12\pi^2M_P^2}\beta.     \label{bound}
    \end{eqnarray}
The fact that these instantons exist only in the finite range of $\varLambda$ implies the restriction of the microcanonical ensemble of universes to this range, which from the viewpoint of string theory can, in particular, be interpreted as the solution of the landscape problem for stringy vacua if one assumes that this model is a low energy approximation of the string theory.

As was recently shown in \cite{hill-top,slih_inflation}, this model with the fundamental cosmological constant can be generalized to the case when the role of an effective $\varLambda$ is simulated by the hill like potential of the inflaton scalar field $\phi$ in the regime of the slow roll approximation, $\varLambda\to V(\phi)/M_P^2$. Then the CFT driven cosmology can be regarded as a source of the new type of hill-top inflation scenario. In particular, it can provide initial conditions for the Starobinsky model of $R^2$-inflation \cite{Starobinskymodel} or the model of Higgs inflation with the Higgs boson playing the role of the inflaton non-minimally coupled to gravity \cite{BezShap,we,RGH}. A major difficulty with this scenario is the problem of hierarchy between the sub-Planckian energy scale of inflation --- the inflaton energy density $V(\phi)\sim 10^{-11}M_P^4$ \cite{RGH} compatible with current CMB observational data \cite{WMAP,Planck} -- and the energy scale of the CFT driven cosmology (\ref{bound}). To match these energy scales one needs the value of $\beta\sim 10^{13}$. In the Standard Model or its GUT generalizations containing only three low spin fields, which can be conformally coupled to gravity, $s=0$, $s=1/2$ and $s=1$, this is of course impossible. Their contribution to $\beta$
    \begin{eqnarray}
    \beta=\frac1{180}\,\big(\mathbb{N}_0+11 \mathbb{N}_{1/2}+
    62 \mathbb{N}_{1}\big)                \label{100}
    \end{eqnarray}
can reach such a magnitude only by the price of unnaturally high numbers of these particle species. Hidden sector of $\mathbb{N}\sim 10^{13}$ low spin particles sounds too unrealistic to be physically acceptable.

Hidden sector of numerous, actually infinitely many, particles is possible in string theory which is believed to underlie the effective quantum field theory and quantum gravity. However, these particles are massive, of the Planckian scale mass, and cannot incorporate local conformal invariance which is a corner stone of the CFT cosmology. On the other hand, there is a growing believe that string theory can be a broken phase of the Vasiliev theory of higher spin gauge fields \cite{Vasiliev}, which was recently very actively considered in various tests of AdS/CFT correspondence including very wide implications of {\em conformal higher spin} (CHS) field models \cite{GKPST,Tseytlin,GiombiKlebanovTseytlin,BBT}. Therefore, it seems reasonable to try as a hidden weakly interacting sector of the CFT cosmology the tower of these CHS fields, that could provide a large value of $\beta$. Though very speculative, in view of problems with unitarity for higher spins, this idea is strongly motivated by the recent observation that the partial contributions $\beta_s$ to (\ref{101}) very rapidly grow with spin as $s^6$ \cite{GKPST,Tseytlin}, so that the needed value can be attained with the finite tower of HSC fields up to $s=100$ containing $\mathbb{N}\sim 10^6$ polarizations \cite{hill-top,slih_inflation}. Therefore, the replacement of multiple species of the same spin by a tower of higher spins is a much more efficient mechanism for large $\beta$, and the goal of this paper is to discuss this mechanism and its possible consequences.

The concept of a hidden sector of CHS fields allows one to solve another important problem in the CFT cosmology. It suggests the mechanism protecting the theory from uncontrollable contribution of perturbatively nonrenormalizable graviton loops. This is achieved by bringing the maximal energy scale of the above cosmological instantons (\ref{bound}) below the gravitational cutoff. With the definition of this cutoff as the scale under which {\em only} the contribution of the {\em graviton} loops (loops containing at least one graviton propagator) is suppressed, there arises a strong distinction between the number of quantum species $\mathbb{N}$ and the CFT central charge $c\sim\beta$ which participate in the expressions for a conventional gravitational cutoff $\varLambda=M_P/\sqrt\mathbb{N}$ \cite{Veneziano,cutoff} and the inflation scale $\varLambda_I\sim M_P/\sqrt\beta$ corresponding to (\ref{bound}). In models of multiple quantum species the cutoff is usually defined as the scale below which the {\em total} one-loop contribution to the graviton propagator is smaller than the tree-level part \cite{Veneziano}, and this leads to the cutoff $\sim 1/\sqrt\beta$ which is not distinguishable from $1/\sqrt\mathbb{N}$ in simple models, but can be very different when $\beta$ and $\mathbb{N}$ are very different in magnitude. Our definition of the cutoff which involves smallness of only the graviton loop corrections leads to the expression $\sim 1/\sqrt\mathbb{N}$ which can be much higher than $1/\sqrt\beta$ for $\beta\gg\mathbb{N}$, and this is exactly the case of CHS fields. In this way we develop the approximation scheme in which the nonrenormalizable graviton sector is subject to effective field theory under the cutoff $M_P/\sqrt\mathbb{N}$, whereas the renormalizable sector of multiple conformal species is treated beyond perturbation theory.

There are several other interesting features associated with the contribution of higher spins to the conformal anomaly (\ref{anomaly}). First, unlike for the case of lower spins $\beta_s$ is negative for fermionic fields with $s\geq 3/2$ \cite{Tseytlin}. This means that overall $\beta$ in CFT cosmology can be negative, and this essentially modifies its scenario. Secondly, irrespective of the sign of $\beta_s$ one can construct the zeta-function regularized sum over infinite set of spins of CHS fields (described by totally symmetric bosonic tensors or fermionic spin-tensors) which yields overall zero value of the total $\beta$ in (\ref{101}) \cite{GKPST,Tseytlin}. This nontrivial manifestation of the AdS/CFT correspondence is conjectured to underlie hypothetical free of trace anomalies -- and therefore quantum consistent -- fundamental theory that might be based on the fusion of string theory and Vasiliev gauge theory of higher spins. Motivated by the prospect of solving the hierarchy problem via CHS fields we consider peculiarities of cosmological scenario driven by their finite tower in case of both positive and negative $\beta$ and by the anomaly free CHS theory with an infinite set of fields.

\section{CFT driven cosmology}
The effective action of the CFT driven cosmology (\ref{eff}) on the FRW metric (\ref{FRW}) with $S^1\times S^3$ topology, $S_{\rm eff}[\,g_{\mu\nu}^{FRW}]\equiv S_{\rm eff}[\,a,N\,]$, was obtained by the conformal transformation to the static Einstein Universe \cite{slih} with a compactified Euclidean time. It consists of the minisuperspace reduced Einstein term, Riegert-Fradkin-Tseytlin action \cite{Riegert} and the contribution of the Einstein static spacetime -- free energy of conformal fields and their vacuum Casimir energy. In units of the rescaled Planck mass $m_P=(3\pi/4G)^{1/2}=(6\pi^2M_P^2)^{1/2}$ it reads \cite{slih}
    \begin{eqnarray}
    &&S_{\rm eff}[\,a,N\,]=m_P^2\int_{S^1} d\tau\,N \left\{-aa'^2
    -a+ \frac\varLambda3 a^3+\,B\left(\frac{a'^2}{a}
    -\frac{a'^4}{6 a}\right)
    +\frac{B}{2a}\,\right\}
    +F(\eta),                           \label{effaction0}
    \end{eqnarray}
where $a'\equiv da/Nd\tau$. The first three terms in curly brackets of (\ref{effaction0}) represent the Einstein action with a fundamental cosmological constant $\varLambda\equiv 3H^2$ ($H$ is the corresponding Hubble parameter). The constant $B$ is a coefficient of the contributions of the conformal anomaly and vacuum (Casimir) energy $(B/2a)$ on a conformally related static Einstein spacetime. It is proportional to the coefficient $\beta$ of the Gauss-Bonnet term $E$ in the trace anomaly of conformal matter fields (\ref{anomaly})
    \begin{eqnarray}
    B=\frac{3\beta}{4 m_P^2}.         \label{B}
    \end{eqnarray}
The free energy $F(\eta)$ of the set of all conformal fields labelled by their spin $s$ also comes from this Einstein space,
    \begin{eqnarray}
    &&F(\eta)=\sum_s\nu_s\sum_{\omega_s}\ln\big(1\mp
    e^{-\omega_s\eta}\big),                 \label{freeenergy}\\
    &&\eta=\int_{S^1} \frac{d\tau N}a.     \label{period}
    \end{eqnarray}
This is a typical boson or fermion sum over field oscillators with energies $\omega_s$ on a unit 3-sphere, $\nu_s$ denoting the number of physical polarizations of a spin-$s$ field, which is negative for fermions, ${\rm sign}\nu_s=\pm1$. The role of the temperature is played here by the inverse of $\eta$ --- an overall circumference of $S^1$ in the $S^1\times S^3$ instanton in units of the conformal time (\ref{period}).

The effective action is independent of the anomaly coefficient $\alpha$, because it is assumed that $\alpha$ is renormalized  to zero by a local counterterm,
    \begin{eqnarray}
    \varGamma_{CFT}\to\varGamma_{CFT}
    +\frac\alpha{384\pi^2}\int d^4x\,g^{1/2}R^2.   \label{alpha_renorm}
    \end{eqnarray}
This guarantees the absence of higher derivative terms in (\ref{effaction0}) \cite{slih,slih_inflation} -- non-ghost nature of the scale factor -- and simultaneously shifts the UV renormalized Casimir energy (which universally expresses in terms of anomaly coefficients $\alpha$ and $\beta$ \cite{universality}) to a partiular value independent of $\alpha$ and proportional to $B/2=\beta/16\pi^2M_P^2$ \cite{slih_inflation,universality},\footnote{This finite renormalization can be generated by the inclusion of the Starobinsky $R^2$-model which also provides the inflaton mode simulating a slowly varying cosmological term which decays at the exit from inflation \cite{hill-top,slih_inflation}.}
    \begin{eqnarray}
    \sum_s\nu_s\sum_{\omega_s}\frac{\omega_s}2=
    \frac38\left(\beta-\frac{\alpha}2\right)\;\to\;
    \frac38\left(\beta-\frac{\alpha}2\right)
    +\frac{3\alpha}{16}
    =\frac38\,\beta.                   \label{Casimir_renorm}
    \end{eqnarray}
Both of these properties are critically important for the instanton solutions of effective equations. The coefficient $\gamma$ of the Weyl tensor term $C^2_{\mu\nu\alpha\beta}$ also does not enter (\ref{effaction0}) because $C_{\mu\nu\alpha\beta}$ identically vanishes for any FRW metric.

The statistical sum (\ref{Z}) is dominated by solutions of the effective equation, $\delta S_{\rm eff}/\delta N(\tau)=0$, which in the gauge $N=1$ reads
    \begin{eqnarray}
    &&-\frac{\dot a^2}{a^2}+\frac{1}{a^2}
    -B \left(\,\frac{\dot a^4}{2a^4}
    -\frac{\dot a^2}{a^4}\right) =
    \frac\varLambda3+\frac{C}{ a^4},\quad
    \dot a=\frac{da}{d\tau},               \label{efeq}\\
    &&C =\frac{B}2+\frac1{m_P^2}\,
    \frac{dF}{d\eta}.                          \label{bootstrap}
    \end{eqnarray}
This is the modification of the Euclidean Friedmann equation by the anomalous $B$-term and the radiation term $C/a^4$. The constant $C$ here characterizes the sum of the Casimir energy and the energy of thermally excited particles with the inverse temperature $\eta$ given by (\ref{period}),
    \begin{eqnarray}
    \frac{dF}{d\eta}=\sum_{s,\omega_s}\frac{|\nu_s|\,\omega_s}
    {e^{\omega_s\eta}\mp 1}.
    \end{eqnarray}
It is a nonlocal functional of the history $a(\tau)$ -- the equation (\ref{bootstrap}) plays the role of the bootstrap equation for the amount of radiation which is determined by the background on top of which this radiation evolves and produces back reaction.

The quadratic equation (\ref{efeq}) can be solved for $\dot a^2$,
    \begin{eqnarray}
    &&\dot{a}^2 = \sqrt{\frac{(a^2-B)^2}{B^2}
    +\frac{2H^2}{B}\,(a_+^2-a^2)(a^2-a_-^2)}
    -\frac{a^2-B}{B},                          \label{mainEq}\\
    &&a_\pm^2\equiv
    \frac{1\pm\sqrt{1-4CH^2}}{2H^2},         \label{apm}
    \end{eqnarray}
where the positive sign of the square root is chosen to provide a periodic oscillation of $a$ between its maximal and minimal values $a_\pm$. In order to guarantee that at $a_-$ is a turning point with a vanishing $\dot a$ the value $a_-^2$ should satisfy the bound $a_-^2>B$. This gives the first two restrictions on the range of $H^2$ and $C$
    \begin{eqnarray}
    H^2\leq\frac1{2B}, \quad C\geq B-B^2H^2,     \label{Hbound}
    \end{eqnarray}
whereas the third one follows from the requirement of real turning points $a_\pm$,
    \begin{eqnarray}
    C\leq \frac1{4H^2}.    \label{upperCbound}
    \end{eqnarray}

There are two sets of solutions of this integro-differential equation \cite{slih,why,DGP/CFT}. The main set consists of periodic $S^3\times S^1$ instantons with the oscillating scale factor -- {\em garlands} that can be regarded as the thermal version of the Hartle-Hawking instantons. The scale factor oscillates $m$ times ($m=1,2,3,...$) between the maximum and minimum values (\ref{apm}), $a_-\leq a(\tau)\leq a_+$, so that the full period of the conformal time (\ref{period}) is the $2m$-multiple of the integral between the two neighboring turning points of $a(\tau)$, $\dot a(\tau_\pm)=0$,
    \begin{eqnarray}
    &&\eta=2m\int_{a_-}^{a_+}
    \frac{da}{\dot{a}a}.                       \label{period1}
    \end{eqnarray}
This value of $\eta$ is finite and determines effective temperature $T=1/\eta$ as a function of $G=3\pi/4m_P^2$ and $\varLambda=3H^2$. This is the artifact of a microcanonical ensemble in cosmology \cite{why} with only two freely specifiable dimensional parameters --- the gravitational and cosmological constants.

According to (\ref{Hbound}) these garland-type instantons exist only in the limited range of the cosmological constant $\varLambda=3H^2$ \cite{slih}. In view of (\ref{Hbound}) and (\ref{upperCbound}) they belong to the domain in the two-dimensional plane of the Hubble constant $H^2$ and the amount of radiation constant $C$. In this domain they form an countable, $m=0,1,2,...$, sequence of one-parameter families -- curves interpolating between the lower straight line boundary $C=B-B^2H^2$ and the upper hyperbolic boundary $C=1/4H^2$. Each curve corresponds to a respective $m$-fold instantons of the above type. Therefore, the range of admissible values of $\varLambda$,
    \begin{eqnarray}
    \varLambda_{\rm min}\leq\varLambda\leq
    \varLambda_{\rm max}=
    \frac{12\pi^2M_P^2}\beta=\frac3{2B},    \label{Lambda_range}
    \end{eqnarray}
has a band structure, each band $\Delta_m$ being a projection of the $m$-th curve to the $H^2$ axis. The sequence of bands of ever narrowing widths with $m\to\infty$ accumulates at the upper bound of this range $H^2_{\rm max}=1/2B$. The lower bound $H^2_{\rm min}$ -- the lowest point of $m=1$ family -- can be obtained numerically for any field content of the model.

Another set of solutions follows from rewriting the effective equation (\ref{efeq}) in the form (retaining in contrast to (\ref{mainEq}) both signs of the square root)
    \begin{eqnarray}
    \dot{a}^2 =1-\frac{a^2}{B}\left(\,1
    \pm\sqrt{1-2BH^2-\frac{B(2C-B)}{a^4}}\,\right) \label{efeq2}
    \end{eqnarray}
and noting that for $C=B/2$ it reduces to \cite{hatch} (without radiation contribution in Lorentzian signature spacetime this solution was derived in \cite{Shapiroetal})
    \begin{eqnarray}
    &&\dot{a}^2 =1-H_\pm^2 a^2,   \label{efeq3}\\
    &&H_\pm^2=\frac{1\pm\sqrt{1-2BH^2}}B=
    \left.\frac1{a_\mp^2}\,\right|_{\,C=B/2}.
    \end{eqnarray}
Obviously the solutions to these two equations, $a(\tau)=\sin(H_\pm\tau)/H_\pm$, represent spherical Euclidean instantons $S_\pm^4$ of the radii $a_\mp$ respectively, or the strings of such spheres touching each other at their poles and forming a ``{\em necklace}" with any number of such spherical beads \cite{hatch}. Note that the value of $C=B/2$ is consistent with the bootstrap equation (\ref{bootstrap}), because the time period for such a necklace consisting of $m$ beads,
    \begin{eqnarray}
    \eta=2m\int_0^{a_\pm}
    \frac{da}{\dot{a}a}=\infty,                       \label{period2}
    \end{eqnarray}
diverges at the poles of spherical beads, where they touch each other -- the range of integration over $a$ in contrast to Eq.(\ref{period1}) is a multiple of the range between $a=0$ at the pole of the 4-sphere $S_\pm^4$ and its value $a_\mp$ at the equator of $S_\pm^4$. Therefore both $F(\eta)$ and $dF(\eta)/d\eta$ vanish and give in view of (\ref{bootstrap}) the value of $C=B/2$.
\begin{figure}[h]
\centerline{\includegraphics[width=7cm]{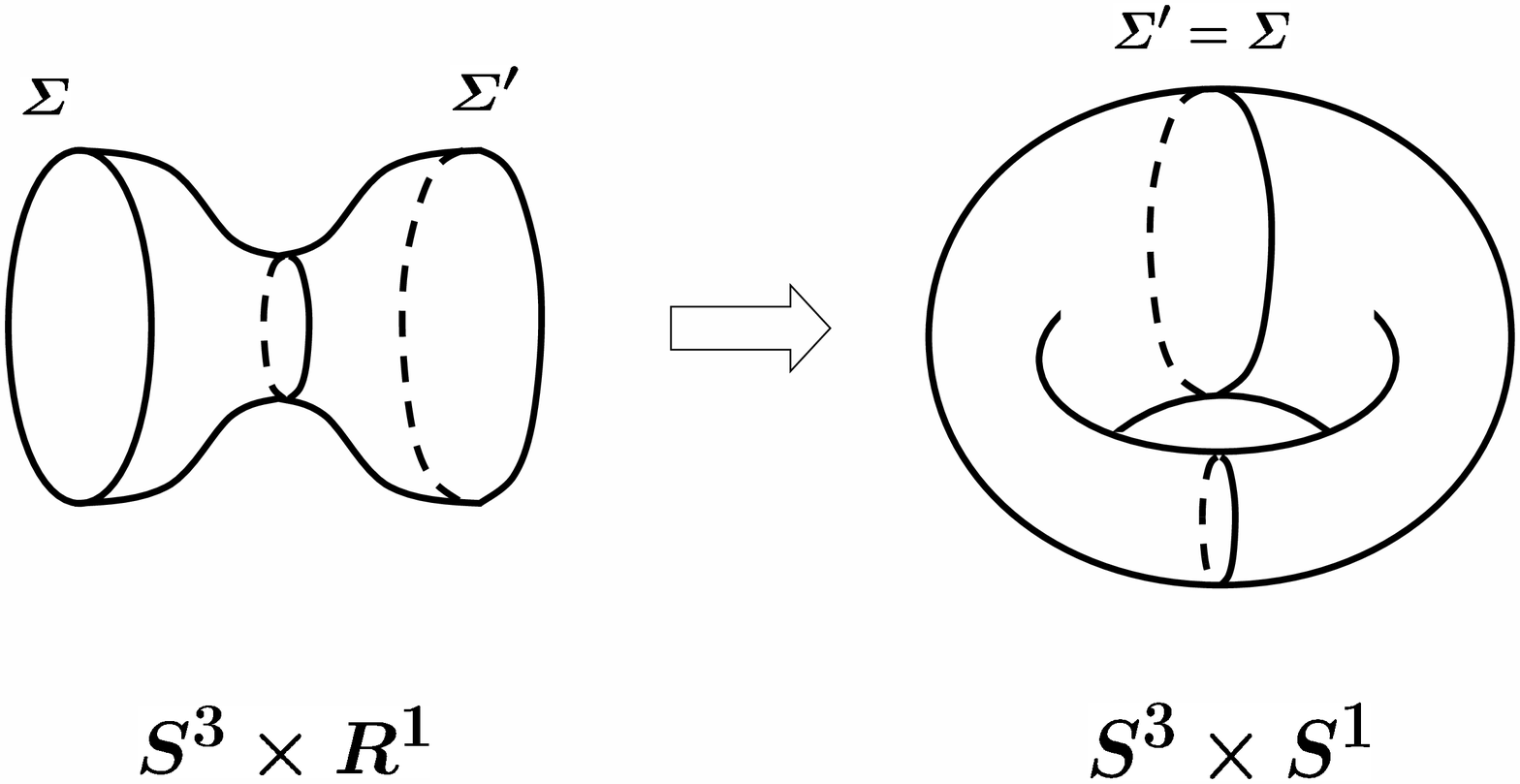}} \caption{\small
Transition from the density matrix to the statistical sum.
 \label{Fig.1}}
\end{figure}

These vacuum (or zero temperature, $1/\eta=0$) necklace instantons existing for all values of $\varLambda=3H^2>0$ are, however, uninteresting because their contribution to the statistical sum is suppressed to zero by their infinite {\em positive} Euclidean action. For $B>0$ the on-shell value of the action (\ref{effaction0}),
    \begin{equation}
    \varGamma_0= F(\eta)-\eta F'(\eta)
    +4m_P^2\int_{S^1}
    \frac{d\tau}{a}\,\dot{a}^2\Big(B-a^2
    -\frac{B\dot{a}^2}{3}\Big)\to+\infty,       \label{action-instanton}
    \end{equation}
diverges to $+\infty$ at the poles of necklace beads with $a=0$, where $|\dot a|=1$ and $B-B\dot a^2/3>0$. Thus the CFT cosmology scenario is free from the infrared catastrophe of vacuum no-boundary instantons, which would otherwise have a {\em negative} tree-level Euclidean action (proportional to $-1/\varLambda\to -\infty$ at $\varLambda\to 0$) and which would imply that the origin of an infinitely big universe is infinitely more probable than that of a finite one. Elimination of this infrared catastrophe is the quantum effect of the trace anomaly which flips the sign of the effective action and sends it to $+\infty$.
\begin{figure}[h]
\centerline{\includegraphics[width=7cm]{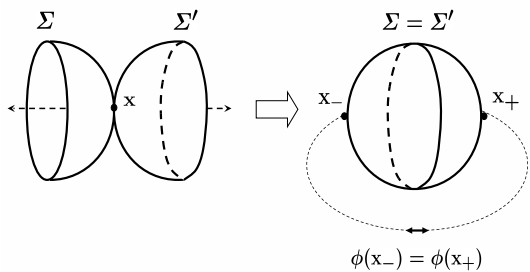}} \caption{\small
Origin of boundary conditions: in the transition to the statistical sum the pinching point $x$ goes over into two different points of $S^4$, $x\to x_\pm$, with equal field values.
 \label{Fig.2}}
\end{figure}

The explanation why the trace anomaly action does not produce the same effect for the no-boundary prescription of Hartle and Hawking consists in the observation \cite{hatch} that the density matrix prescription in the CFT cosmology, despite the same $S^4$-geometry of the cosmological instanton, has boundary conditions other than those of the no-boundary case. Graphical representation of the origin of the periodic garland instanton from the density matrix, whose two arguments are associated with spatial hypersurfaces $\varSigma$ and $\varSigma'$, is demonstrated on Fig.1 for a single-fold case. Analogous single-bead necklace instanton originates by a similar procedure shown on Fig.2. It implies that prior to transition to the statistical sum the values of the fields on two hemispheres should coincide at the point where these hemispheres touch. After tracing out the fields at the identified surfaces $\varSigma'=\varSigma$ the hemispheres get glued into a complete $S^4$ instanton, but its antipodal pole points should still be identified as well as their field values $\phi(x_-)=\phi(x_+)$. This is the additional boundary condition which is absent in the conventional no-boundary prescription. And, as shown in \cite{hatch}, path integration over the quantum field on a generic Euclidean manifold with the identification of values of this field at two spacetime points leads to the suppression of the result to zero by its one-loop prefactor. This result strongly relies on positivity of the classical Euclidean action or unitarity, which is apparently related to the positivity of $B\sim\beta$ in the above derivation.

\section{CFT driven cosmology with CHS fields }

The main motivation for introducing CHS fields in cosmology is an attempt of solving the hierarchy problem. Realistic CFT driven cosmology capable of generating a finite inflationary stage arises by the generalization of the model (\ref{tree0}) to the case of effective cosmological term with a slowly varying $\varLambda$. $\varLambda$ is simulated by a potential $V(\phi)$ of a dynamical scalar field $\phi$ -- the inflaton in the regime of the slow roll from the slope of $V(\phi)$. Such a model of a new hill-top inflation scenario was recently built in \cite{hill-top,slih_inflation}. It turns out that the slow roll conditions -- smallness of inflationary parameters $\epsilon$ and $\eta$, which determine the properties of the primordial CMB power spectrum \cite{ChibisovMukhanov}, are satisfied for those models which are very close to the upper boundary of the effective cosmological constant range (\ref{Lambda_range}). This leads to a major difficulty in a realistic inflationary scenario -- the problem of hierarchy between the Planckian scale of this bound and the estimates for the scale of inflation based on the CMB data in the models which provide a good fit of these data \cite{WMAP,Planck}. Here we show that using CHS fields suggests a solution to this hierarchy problem and, moreover, provides a mechanism protecting the obtained results from the contribution of nonrenormalizable graviton loops.

\subsection{Hierarchy problem}

Two models which perhaps give the best fit are the Starobinsky $R^2$-inflation model and quantitatevely very close to it Higgs inflation model with a large non-minimal coupling of the Higgs inflaton to curvature \cite{BezShap,we,RGH}. In particular, the latter model establishes a very convincing relation between the observable CMB spectral parameter $n_s\simeq 0.96$ and the value of the Higgs mass very close to the one discovered at LHC \cite{we,RGH}.

For Higgs inflation with a large non-minimal coupling the energy density (in the Einstein frame of fields) at the start of inflation should be essentially sub-Planckian, $V\sim 10^{-11}M_P^4$ \cite{we}. Matching with the upper bound (\ref{Lambda_range})
    \begin{eqnarray}
    M_P^2\varLambda\sim \frac{3M_P^2}{2B}
    =\frac{12\pi^2}\beta\, M^4_P.
    \end{eqnarray}
implies that the  total beta must be of the order of magnitude
    \begin{eqnarray}
    \beta\sim 10^{13}.     \label{beta_bound}
    \end{eqnarray}
In order to reach this value with the conventional low spin particle phenomenology characteristic of the Standard Model one would need unrealistically high numbers of conformal invariant scalar bosons $\mathbb{N}_0$, Dirac fermions $\mathbb{N}_{1/2}$ and vector bosons $\mathbb{N}_{1}$ in the expression  (\ref{100}) for $\beta$.

Hidden sector of so numerous low spin weakly interacting particles does not seem to be realistic. However, unification of interactions inspired by the ideas of string theory, holographic duality \cite{AdS/CFT0} and higher spin gauge theory \cite{Vasiliev} suggests that this hidden sector might contain conformal higher spin (CHS) fields described by totally symmetric tensors and spin-tensors \cite{GKPST,Tseytlin,GiombiKlebanovTseytlin}, and  the total value of $\beta$ consists of the additive sum of all their partial contributions (\ref{101}). Recently there was essential progress in the theory of these fields. In particular, it was advocated in \cite{GKPST,Tseytlin} that the values of $\beta_s$ can be explicitly calculated for both bosons and Dirac fermions of a generic spin $s$. They read
    \begin{eqnarray}
    &&\beta_s=\frac1{360}\,\nu_s^2(3+14\nu_s),
    \quad \nu_s=s(s+1),\quad s=1,2,3,...\,,       \label{boson}\\
    &&\beta_s=\frac1{720}\,
    \nu_s(12+45\nu_s+14\nu_s^2),\quad
    \nu_s=-2\Big(s+\frac12\Big)^2,\quad
    s=\frac12,\frac32,\frac52,...\,,            \label{fermion}
    \end{eqnarray}
where $\nu_s$ is their respective number of dynamical degrees of freedom -- polarizations (negative for fermions).\footnote{Spin zero field -- a scalar conformally coupled to gravity -- for certain group-theoretical reasons \cite{Tseytlin} is not included in the list of bosonic CHS, and its value $\beta_0=1/180$ violates the rule (\ref{boson}) which is valid for $s>0$.} Though these fields serve now basically as a playground for holographic AdS/CFT duality issues and suffer from the problems of perturbative unitarity, which is anticipated to be restored only at the non-perturbative level,\footnote{Lack of perturbative unitarity is associated with the fact that CHS fields are not free from ghosts -- their kinetic operator contains higher derivatives of order $2s$.} it is worth trying to exploit them as a possible solution of hierarchy problem in the CFT driven cosmology.

Strong motivation for this is that partial contributions of individual higher spins rapidly grow with the spin as $s^6$, so that the tower of spins up to some large $S$ generates the total value of $\beta$
    \begin{eqnarray}
    \beta\simeq \frac7{180}\int_0^S ds\,s^6=\frac{S^7}{180}  \label{beta}
    \end{eqnarray}
(for simplicity we consider only bosons and assume that every higher spin species is taken only once, $\mathbb{N}_s=1$). Therefore, in order to provide the hierarchy bound (\ref{beta_bound}) the maximal spin should be $S\sim 100$, which corresponds to the following estimate of the total number of particle modes (polarizations) in the hidden sector of the theory
    \begin{eqnarray}
    \mathbb{N}=\sum_s\nu_s\simeq\int_0^S ds\,s^2\sim 10^6. \label{nu}
    \end{eqnarray}
When, instead of a tower of spins, the cosmological model is driven by a conformal field of an individual spin $s\gg 1$ with $\beta_s\simeq s^6/25$ and $\nu_s\simeq s^2$, then the needed value of spin and the number of polarizations are
    \begin{eqnarray}
    s\simeq 200,\quad \mathbb{N}_s\equiv\nu_s\simeq 4\times 10^4. \label{ind_spin}
    \end{eqnarray}

Quite interestingly, in the case of (\ref{beta})-(\ref{nu}) this coincides with the estimate for the average value of $\beta$ per one conformal degree of freedom, $\tilde\beta\equiv\beta_{\rm boson}/\mathbb{N}_{\rm boson}\sim 10^6$  at which the thermal correction to the spectral parameter of CMB, $\Delta n_s^{\rm thermal}\sim -0.001$, depending on the properties of the hill like inflaton potential might appear in its third decimal order \cite{CMBA-theorem,slih_inflation} -- the precision anticipated to be reachable in the next generation of CMB observations following Planck. For the case of the individual spin (\ref{ind_spin}) the thermal correction is even stronger and can appear in the second or even the first decimal order. This means that a potential resolution of the hierarchy problem in the CFT scenario via CHS simultaneously would make measurable the thermal contribution to the CMB red tilt, which is complementary to the conventional tilt caused by the deviation of the slow roll evolution from the exact de Sitter evolution \cite{ChibisovMukhanov}.

\subsection{Stability of quantum corrections and gravitational cutoff}
CHS fields provide a mechanism which does not only solve the phenomenological problem of hierarchy between the Planckian and inflation scales, but is also likely to justify the approximation underlying the predictions of the above type. As mentioned in Introduction this approximation goes beyond semiclassical expansion, because its subleading one-loop order is critically important for the construction of the model. In particular, it contributes a large one-loop contribution whose balance against the tree-level part establishes the upper bound (\ref{bound}) on the energy scale of inflation. Therefore, the inflation scale
    \begin{equation}
    \varLambda_I=\frac{M_P}\beta    \label{LambdaI}
    \end{equation}
is deeply below the Planck scale for a large $\beta$ generated by large numbers of quantum species $\mathbb{N}_s$,$M_P/\sqrt\beta\sim M_P/\sqrt\mathbb{N}$, $\mathbb{N}=\sum_s \mathbb{N}_s$. This seems to imply validity of the semiclassical expansion in which all graviton loops can be disregarded. However, there is a problem associated with the fact that quantum gravity with a large number $\mathbb{N}$ of quantum species has the effective field theory cutoff
    \begin{equation}
    \varLambda=\frac{M_P}{\sqrt\mathbb{N}},    \label{cutoff}
    \end{equation}
which decreases for growing $\mathbb{N}$ along with the scale (\ref{LambdaI}) and can be even higher than the latter. This is a well known statement based on perturbation theory arguments \cite{Veneziano} or implications of the Hawking radiation from a semiclassical black hole \cite{cutoff}. Within a conventional perturbation theory this completely breaks the predictions in our model, because we cannot guarantee smallness of quantum corrections due to nonrenormalizable graviton loops -- the theory is not protected from uncontrollable radiative corrections.

The origin of the gravitational cutoff (\ref{cutoff}), below which a nonrenormalizable contribution of graviton loops can be disregarded, is based on the following reasoning. Suppose we have $\mathbb{N}$ quantum species which generate in the external gravitational field a full set of Feynman diagrams. Inclusion of quantum graviton loops consists in the insertion into these diagrams graviton propagators with the relevant vertices of gravity-matter interaction. Each graviton propagator carries extra factor of $1/M_P^2$ which suppresses a relevant background quantity of the scale $\varLambda^2$ -- a spacetime curvature or a second order spacetime derivative. With a small $O(1)$ number of quantum fields the suppression factor for this contribution would be $\varLambda^2/M_P^2$ which implies a standard Planck scale cutoff $\varLambda=M_P$. However, with many quantum fields this contribution gets enhanced.

For noninteracting multiple species the above contribution is multiplied by the first power of $\mathbb{N}$, since the graviton propagator connects two vertices of one and the same field. This is because another matter field propagator, connecting these vertices, for free (linear) fields can connect only the vertices of one and the same species, whereas another graviton propagator which could have connected vertices of different fields would give (together with the first added propagator) a subdominant contribution $\sim 1/M_P^4$. Therefore, the overall suppression factor for the insertion of any single graviton propagator becomes $\mathbb{N}\varLambda^2/M_P^2$, and the cutoff equals (\ref{cutoff}).\footnote{For interacting quantum species the inserted graviton propagator can connect vertices of different fields (the full $\mathbb{N}\times\mathbb{N}$ matrix propagator of quantum species can be non-diagonal), so that the suppression factor equals $\mathbb{N}^2\varLambda^2/M_P^2$ and the cutoff seemingly reduces to $\varLambda_{\rm int}=M_P/\mathbb{N}$. But off-diagonal elements should carry the coupling constant which should scale as $1/\mathbb{N}$ to guarantee perturbative regime in the matter field sector, and this raises $\varLambda_{\rm int}$ back to the non-interacting case of (\ref{cutoff}). I am grateful to S.Sibiryakov for the discussion of this point.} Thus, a hidden sector of multiple quantum species does not help to protect the model from unrenormalizable graviton corrections. Its energy scale $M_P/\sqrt\beta$ with $\beta\sim\mathbb{N}$ either coincides with the gravitational cutoff (\ref{cutoff}) or even exceeds it. This is the consequence of the universality of the gravitational interaction -- no matter how small is the coupling of the hidden sector to observable sector, universal gravitational interaction cannot be reduced below the gravitational cutoff.

It should be emphasized here that in the above derivation of the gravitational cutoff we did not require smallness of the {\em full} quantum (say one-loop) correction relative to the tree-level part as it was done in \cite{Veneziano}. In the perturbation theory approach of \cite{Veneziano} this would have led instead of $\mathbb{N}$ to the central charge of the theory -- the relevant coefficient of the conformal anomaly $c\sim\beta$, and the cutoff would be $M_P/\sqrt\beta$. Rather, in our definition of the cutoff we demanded smallness of only the graviton corrections -- all nonrenormalizable contributions containing any single graviton propagator. On the contrary, the contribution of purely quantum species loops can be large and treated beyond perturbation theory, as it was done in the construction of cosmological instantons above. With this definition the cutoff is inverse proportional to $\sqrt\mathbb{N}$ rather than $\sqrt\beta$. For conventional low spin theories these two quantities are of the same order of magnitude and are usually not distinguished from one another in the cutoff expression.

Remarkable property of CHS fields is that they qualitatively change this situation. This is because for this theory -- either for an individual sufficiently high spin $s$ or for a tower of spins up to some $s=S\gg 1$ -- the parameter $\beta$ grows with the spin much faster than the number of fields. For an individual spin $s$ the role of the number of quantum species is played now by the number of polarizations $\nu_s\simeq s^2$, while for a tower of spins (for bosonic case with every spin taken once) it is
    \begin{equation}
    \mathbb{N}=\sum\limits_s \nu_s\simeq \frac{S^3}3.
    \end{equation}
Correspondingly, the total value of $\beta$ equals $\beta_s\simeq 7s^6/180$ for an individual spin and $\beta\simeq S^7/180$, see Eq.(\ref{beta}), for such a tower of spins. Therefore the ratio of the inflation scale (\ref{LambdaI}) to the cutoff (\ref{cutoff}) is decreasing with the growth of the individual spin $s$ or the height of the spin tower $S$ respectively as\footnote{Here and above we disregarded powers of $2\pi$ which should equally enter the expressions for the inflation scale (\ref{LambdaI}), cf. Eq.(\ref{bound}), and the gravitational cutoff (\ref{cutoff}) and, therefore, cancel out in their ratio.}
    \begin{eqnarray}
    &&\frac{\varLambda_{I,s}}{\varLambda_{s}}
    =\sqrt{\frac{\mathbb{N}_s}{\beta_s}}\simeq\frac5{s^2},\\
    &&\frac{\varLambda_{I,S}}{\varLambda_{S}}
    =\sqrt{\frac{\mathbb{N}}{\beta}}\simeq\frac{\sqrt{60}}{S^2},
    \end{eqnarray}
where the inflation scales and cutoffs labeled by $s$ and $S$ obviously denote the cases of the individual spin or a tower of those. In both cases of the individual spin (\ref{ind_spin}), $s\sim 200$, and the tower of spins (\ref{beta})-(\ref{nu}), $S\sim 100$,  these ratios are very small and range within the limits $10^{-4}\div 10^{-3}$. Therefore, the model is in the quantum state three or four orders of magnitude below its gravitational cutoff, and an uncontrollable contribution of nonrenormalizable graviton loops is negligible.

For strongly coupled species with the cutoff $\varLambda_{\rm int}=M_P/\mathbb{N}$ mentioned above (see footnote 4) these ratios are higher $\varLambda_{I,s}/\varLambda_{{\rm int},s}\simeq 5/s$, $\varLambda_{I,S}/\varLambda_{{\rm int},S}\simeq\sqrt{20/S}$, and amount to $0.03\div 0.4$. So for a tower of spins, $S\sim 100$, the system becomes too close to the gravitational cutoff. This case is, however, not so important because (non-diagonal) propagator mixing between the CHS species is possible only on the nonzero background of these fields, whereas we assume that in conformal cosmology it is vanishing -- conformal species have vanishing expectation values and contribute only via their quantum fluctuations.

Thus the model is deeply below the gravitational cutoff, and we remain with the dominant quantum contribution of only the CHS fields in the external gravitational field. This contribution is big, because it is weighted by $\beta\gg\mathbb{N}$, and it is treated beyond perturbation theory. For linear fields it is exhausted by the one-loop order, which is exactly calculable for the FRW metric by the trace anomaly method described above. The generalization to a nonlinear case also looks straightforward. It is important that this sector is perturbatively renormalizable, which is obvious for low spin fields $s=0,1/2,1$, while for higher order spins it directly follows from their higher-derivative nature -- the conformal field of spin $s$ has a kinetic operator of order $2s$ \cite{Tseytlin}.  Due to renormalizability gravitational multi-loop counterterms are exhausted by the same three tensor invariants as in the one-loop order, $C_{\mu\nu\alpha\beta}^2$, $E$ and $\Box R$ ($R^2$ is absent due to conformal invariance) which generate the three-parameter trace anomaly (\ref{anomaly}), and the principal effect of nonlinearity is a slow logarithmic RG running of $\alpha$, $\beta$ and $\gamma$ \cite{Komargodski}. Moreover, for the FRW background this sector is actually free from {\em logarithmic} UV divergences, because on this closed $S^1\times S^3$ instanton the counterterms -- $\int d^4x\,g^{1/2} C_{\mu\nu\alpha\beta}^2$, the Euler number (contributed by the Gauss-Bonnet invariant $E$) as well as $\int d^4x\,g^{1/2}\Box R$  -- are all vanishing. Thus, this sector of the model is free from UV divergences, and its power divergences are absorbed by the renormalization of the cosmological $\varLambda$ and gravitational $M_P^2$ coupling constants. It is in terms of these two {\em renormalized} constants and without any renormalization ambiguity that the FRW cosmological instantons were built within the dynamically suppressed energy scale (\ref{bound}).

Note that the $R^2$-term of the Starobinsky model, considered above (\ref{alpha_renorm}) and used for a {\em finite} renormalization of the Casimir energy (\ref{Casimir_renorm}) and simulation of the effective cosmological term \cite{hill-top}, belongs to UV finite sector. Therefore, elimination of higher-derivative ghosts in the gravitational sector by this finite renormalization cannot be broken by leading radiative corrections, which justifies the criterion of ``naturalness" in this model.

\section{Peculiarities of fermionic and anomaly-free CHS models}
Eq.(\ref{fermion}) shows that when the model is dominated by CHS fermions the total $\beta$ and $B$ can be negative, because fermionic $\beta_s$ are negative starting with $s=3/2$. Apparently, this is associated with the fact that HSC fields are not free from ghosts -- their kinetic operator contains higher derivatives of order $2s$. There exist a hope that at the nonperturbative level CHS theories can be rendered unitarity, so that there effect can still be interesting within the CFT cosmology scenario. The picture of this scenario with a negative $B$ is somewhat different from the case of $B>0$ and looks as follows.

First consider the garland instantons. For a negative $B=-|B|$ the equation (\ref{mainEq}) should be chosen in the form with the negative sign of the square root
    \begin{equation}
    \dot{a}^2 = 1+\frac{a^2}{|B|}
    -\sqrt{\Big(1+\frac{a^2}{|B|}\Big)^2
    -\frac{2H^2}{|B|}\,(a_+^2-a^2)(a^2-a_-^2)},\label{negativeB}
    \end{equation}
if we want periodic oscillations between $a_-$ and $a_+$. Here the argument of the square root equals $a^4(1+2|B|H^2)/B^2+1+2C/|B|$ and is positive definite, and also no restrictions on the range of $H^2$ and $C$ similar to (\ref{Hbound}) follow. Apriori, in the two dimensional plane of $H^2>0$ and $C>0$ garland instanton solutions can occupy all points below the hyperbola $C=1/4H^2$. However, the bootstrap equation (\ref{bootstrap}) with a negative $B$ imposes the upper bound on the value of the conformal time period of garland instantons, and this leads to a reduction of this domain. The constant $C$ cannot be negative because $a_-^2$ should be positive for garlands, so that the range $0<C<1/4H^2$ gives, in view of $dF/d\eta$ being a monotonically decreasing function of $\eta$, a restriction on the domain of possible $\eta$
    \begin{eqnarray}
    &&\frac{|B|}2<\frac1{m_P^2}\,\frac{dF}{d\eta}<\frac1{4H^2}+\frac{|B|}2,\quad
    \eta_{\rm max}>\eta>\eta_{\rm min}.
    \end{eqnarray}
This immediately rules out solutions belonging to $H^2$ and $C$ axes of the curvilinear triangle, because both for $C=0$ ($a_-=0$) and $H^2=0$ ($a_+^2\to\infty$) the time period is infinite, which contradicts $\eta\leq\eta_{\rm max}$. Similarly, the asymptotics $H^2\to 0$ cannot be reached along the upper hyperbolic boundary, $C=1/4H^2$, because the bootstrap equation would imply that $dF/d\eta\sim m_P^2C= m_P^2/4H^2\to\infty$ and $\eta\to 0$ (in view of bosonic contributions to $dF/d\eta$ tending to infinity for vanishing $\eta$). But this is impossible because at this boundary the conformal time period is exactly calculable\footnote{At $C\to1/4H^2$ the turning points $a_\pm$ approach each other and the integral for $\eta$ becomes exactly calculable \cite{slih}.} \cite{slih} and equals
    \begin{equation}
    \eta\,\big|_{\,C=1/4H^2}=\pi\sqrt{2(1-2BH^2)}\geq\pi\sqrt2.
    \end{equation}
The values of $H^2$ are also bounded from above because for $H^2\to\infty$ the conformal time period $\eta\sim H\sqrt{|B}|\to\infty$, and large $H$ are also ruled out by the maximal value of $\eta$. In contrast to the case of positive $B$, however, the upper bound on the cosmological constant is not universal and can be obtained only numerically.

Thus, negative $B$ rules out infrared catastrophe in the distribution of garland type cosmological models with $H^2\to 0$, because there are no corresponding garland instantons in this limit. This restricts the range of the cosmological constant from above. One would think that the upper limit on $\eta$ might also bound the value of $H^2$ and solve the hierarchy problem. Say, on the hyperbolic boundary with $\eta =\pi\sqrt{2(1+2|B|H^2)}$ it reads $1+2|B|H^2\leq \eta_{\rm max}^2/2\pi^2$. But this bound turns out to be to weak to provide $H^2\simeq 10^{-12}m_P^2$, unless the value of $m_P^2|B|$ in $H^2/m_P^2\leq\left(\eta^2_{\rm max}/2\pi^2-1\right)/{2|B|m_P^2}\sim 10^{-12}$ is very big, which again implies a high tower of fermionic CHS.

The domain of negative $C$ and $B$ in (\ref{bootstrap}) corresponds to {\em necklace} instantons described above in Sect.2. For $C<0$ the value of $a_-^2$ is also negative, so that a sensible minimal value of the scale factor is actually zero. This does not lead to a contradiction with (\ref{negativeB}), because the conformal time $\eta$ in view of Eq.(\ref{period2}) diverges to infinity at the lower limit and $C=-|B|/2$, so that the geometry smoothly closes at $a=0$ without a conical singularity to form a round sphere $S^4_-$ (or a necklace -- the string of round spheres) for all positive $H^2$. However, in contrast to necklace instantons with a positive $B$ they are not ruled out by their on-shell effective action (\ref{action-instanton}) because $B<0$ and the action is infinitely negative $\varGamma_0=-\infty$. Note that this divergence occurs not at a particular value of $H^2$ -- the statistical sum is badly defined for all values of $H^2$, and the situation is qualitatively different from the no-boundary state. Therefore, the cosmology with $\beta<0$ is in principle inconsistent, because its full statistical sum contains this divergent contribution of necklace instantons. The additional ensemble of garland instantons, discussed above, contributes to the statistical sum a finite part, but it is hard to interpret it on top of the divergent part of the necklace distribution.

Another interesting case of CHS theory is the infinitely high tower of spins
$S\to\infty$. In this case the total $\beta$ and other characteristics of the CHS tower are formally divergent, but in the case when all spin particles are taken only once, $\mathbb{N}_s=1$, the use of $\zeta$-function regularization for the divergent spin series gives a remarkable result -- an overall $B$ becomes zero \cite{GKPST,Tseytlin},
    \begin{eqnarray}
    \sum\limits_s\beta_s=0, \quad B=0.
    \end{eqnarray}
This result holds independently for boson, $s\geq 1$, and fermion, $s\geq1/2$ \cite{Tseytlin}, CHS towers  and underlies a non-trivial check of the $AdS_5/CFT_4$-correspondence in which the vanishing result on the 5-dimensional AdS side simply follows from the absence of conformal anomaly in odd-dimensional spacetime.

Explicit summation over spins is possible not only for trace anomaly coefficients but also for the thermal free energies of CHS particles \cite{BBT}. This can be done by expanding the logarithm in (\ref{freeenergy}) in the sum over ``one-particle" statistical sums ${\cal Z}_s(m\eta)$ with growing inverse temperature $m\eta$ and interchanging the order of summation over spins and equidistant levels of the inverse temperature,
    \begin{eqnarray}
    &&F_{\rm boson}(\eta)=
    -\sum\limits_{m=1}^\infty\frac1m {\cal Z}_{\rm boson}(m\eta),\\
    &&{\cal Z}_{\rm boson}(\eta)=\sum\limits_{s=1}^\infty{\cal Z}_s(\eta),\quad
    {\cal Z}_s(\eta)=\sum\limits_{\omega_s}|\nu_s|\,e^{-\eta\omega_s}.
    \end{eqnarray}
Explicit summation of ${\cal Z}_s(\eta)$ over spin $s$ gives the answer \cite{BBT}
    \begin{eqnarray}
    &&{\cal Z}_{\rm boson}(\eta)=
    -\frac{e^{-2\eta}(11+26 e^{-\eta}+11 e^{-2\eta})}{6\,(1-e^{-\eta})^6}\simeq
    -\frac{11}6\,e^{-2\eta},\\
    &&F_{\rm boson}(\eta)\simeq
    -{\cal Z}_{\rm boson}(\eta)\simeq
    \frac{11}6\,e^{-2\eta},
    \quad \eta\to\infty,                     \label{bosonF}
    \end{eqnarray}
which gives a good approximation for the sum over $m$ in the limit of large $\eta$. Irrespective of this approximation, the answer for the total free energy, given by this convergent sum over $m$, has important property -- unexpectedly it is negative. Similarly for fermions
    \begin{eqnarray}
    &&F_{\rm fermion}(\eta)=
    \sum\limits_{m=1}^\infty\frac{(-1)^m}m {\cal Z}_{\rm fermion}(m\eta),\\
    &&{\cal Z}_{\rm fermion}(\eta)=\sum\limits_{s=1/2}^\infty{\cal Z}_s(\eta)\nonumber\\
    &&\qquad\qquad\quad=-\frac{e^{-3\eta/2}(1+23 e^{-\eta}+23 e^{-2\eta}+e^{-3\eta})}{3\,(1-e^{-\eta})^6}\simeq -\frac13\,e^{-3\eta/2},
    \quad \eta\to\infty.    \label{fermionF}
    \end{eqnarray}
Therefore, both bosonic and fermionic theories of CHS have negative thermal energy densities $dF/d\eta$.

Obviously, this unnatural conclusion is the result of the oversubtraction performed by the $\zeta$-function renormalization of \cite{GKPST,Tseytlin,GiombiKlebanovTseytlin,BBT} which annihilates all UV divergences in the infinite sum over spins. This is clearly seen in the special cutoff regularization by the parameter $\epsilon\to 0$
    \begin{eqnarray}
    &&{\cal Z}_{\rm boson}^\epsilon(\eta)=
    \sum\limits_{s=0}^\infty
    e^{-\epsilon(s+1/2)}{\cal Z}_s(\eta),   \label{reg}\\
    &&{\cal Z}_{\rm boson}^\epsilon(\eta)=
    \frac{4e^{-2\eta}}{(1-e^{-\eta})^4\epsilon^2}
    -\frac{e^{-2\eta}(11+26 e^{-\eta}
    +11 e^{-2\eta})}{6\,(1-e^{-\eta})^6}
    +O(\epsilon).                  \label{bosonF1}
    \end{eqnarray}
This regularization retains power divergences which give a dominant positive contribution, while the $\zeta$-function regularization subtracts these divergences and leaves us with the negative remnant. Potential justification for this property could be a conjectured nonperturbative treatment of such theories which solves the problem of their unitarity. We will see now that outside of this nonperturbative approach these models recover the situation of the vacuum no-boundary instantons with the infrared catastrophe of $\varLambda\to 0$.

Indeed, for $B=0$ and $dF/d\eta<0$ eqs. (\ref{efeq})-(\ref{bootstrap}) lead to a negative $C<0$, so that the turning points (\ref{apm}) cannot be both positive. Similarly to the situation a negative $C$ above, the negative value of $a_-^2$ implies that the minimal value of the scale factor is zero, which does not lead to a conical singularity of the solution at $a_-=0$, because the conformal time $\eta$ in view of (\ref{period1}) diverges to infinity and $C\to 0$. The solution becomes $a^2(\tau)=\sin^2(H\tau)/H^2$, which represents the Hartle-Hawking instanton $S^4$. In view of the overall $B$ equal to zero, the on-shell action (\ref{action-instanton}) does not diverge -- its integral term is finite for all $H^2$. It coincides with the tree-level action of the no-boundary instanton -- the 4-sphere of the radius $1/H$ -- and diverges to $-\infty$ at $H^2\to 0$, $\varGamma_0=-2m_P^2/3H^2$.

\section{Conclusions}
Our main conclusion is that cosmology with numerous CHS fields can solve the problem of hierarchy between the Planck scale and the subplanckian inflationary scale compatible with CMB observations within a number of inflation models including $R^2$-gravity and Higgs inflation model \cite{hill-top,slih_inflation}. For that a finite height tower of bosonic and fermionic CHS fields should generate a large positive coefficient $\beta$ of the Gauss-Bonnet invariant in the overall conformal anomaly of the theory.

In CHS models the coefficient $\beta$ is much higher than the number of field polarizations $\mathbb{N}$ which means that the energy scale of the model $M_P/\sqrt\beta$ is essentially (by three or four orders of magnitude for $R^2$-gravity and Higgs inflation model) below the gravitational cutoff $M_P/\sqrt\mathbb{N}$, the latter being determined from the requirement of smallness of graviton loop corrections. This justifies a special approximation scheme in which these nonrenormalizable corrections are treated within the effective field theory below this cutoff, while the dominant quantum contribution of CHS fields is treated beyond perturbation theory and dynamically puts the bound on the energy scale of the model.

Though it fits presently very popular ideas of string theory, higher spin gauge theory \cite{Vasiliev} and holographic duality \cite{GKPST}, this mechanism of a large $\beta$ is still vulnerable to criticism regarding the problem of perturbative unitarity for CHS fields due to their higher-derivative nature. How important is the lack of unitarity in the hidden sector of our model? Free or weakly selfinteracting CHS fields are coupled to the observable sector only gravitationally, but below the gravitational cutoff this coupling is apparently suppressed similarly to the contribution of graviton loops and the loops containing any single graviton propagator. This effective decoupling essentially reduces the effects of non-unitarity. Of course, this is not a fundamental solution of the problem, which is expected to be achieved only at the nonperturbative level under a better understanding of the gauge theory of interacting higher spin fields.

The problem of unitarity can manifest itself and even can be destructive for a negative $\beta$, which is possible in the case of the dominant contribution of higher spin conformal fermions with $s\geq 3/2$. As we saw, cosmology with $\beta<0$ is inconsistent, because its statistical sum contains a divergent contribution of necklace instantons. The source of this inconsistency is apparently related to the breakdown of perturbative unitarity. Suppression of vacuum necklace instantons, mentioned in Sect.2, is based  on the Gaussian integration with a positive definite quadratic form \cite{hatch}. The boundary condition $\phi(x_+)=\phi(x_-)$, induced by the microcanonical density matrix as depicted in Fig.2, can be enforced with the Lagrange multiplier $\lambda$ in the path integral for a generic theory with the quadratic action $S[\,\phi\,]=\frac12\int dx\,\phi F\phi$. Therefore, the modification due to this boundary condition reduces to a single Gaussian integral over $\lambda$
    \begin{eqnarray}
    &&Z=
    \int D\phi\;d\lambda\,\exp\Big(-S[\,\phi\,]
    +i\lambda\big(\phi(x_+)-\phi(x_-)\big)\Big)\nonumber\\
    &&\qquad\qquad=\big({\rm Det} \,F\big)^{-1/2}\int d\lambda\,\exp\left(-\frac12\int dx\,dy\,J_\lambda(x)G(x,y)
    J_\lambda(y)\right)            \label{1000}
    \end{eqnarray}
with the local source $J_\lambda(x)$ peaked at $x_\pm$, $J_\lambda(x)\equiv \lambda\big(\delta(x-x_+)-\delta(x-x_-)\big)$, and the kernel $G(x,y)$ -- the Green's function of $F$. Therefore, the statistical sum reduces to zero,  $Z\sim(G(x_+,x_+)+G(x_-,x_-)-2G(x_+,x_-))^{-1/2}=0$, in view of the divergent coincidence limits for $G(x,y)$, $G(x_+,x_+)=G(x_-,x_-)=\infty$. Violation of unitarity is associated with the indefiniteness of the operator $F$ which makes $Z$ divergent, and this is consistent with the divergence to $-\infty$ of the integral part of (\ref{action-instanton}) for $B<0$.

The situation with the anomaly free theory of the infinite chain of CHS fields is different. Its statistical sum is finite, but it suffers from the infrared catastrophe at $\varLambda\to 0$ similar to the problem with the no-boundary prescription. Therefore, it is hardly acceptable as a source of cosmological initial conditions, because it suggests that the origin of infinitely big universe is infinitely more probable than that of the finite one. This means that the anomaly free CHS model, as a potential basis of a consistent theory ``of everything", does not give any advantages in context of cosmological applications. What ruins these advantages is the oversubtraction of power divergences by $\zeta$-regularization of the sum over spins in (\ref{bosonF1}), which leads to a negative value of the thermal energy -- the other side of breakdown of perturbative unitarity. Apparently, the disregard of power divergences in the sum over spins physically is not such a harmless procedure. Their recovery in the regularization (\ref{reg}) with the parameter $\epsilon\to 0$, which is equivalent to the cutoff in the height of the spin tower at $S\sim 1/\epsilon$, brings us back to the predictions of the model with a finite tower of CHS fields. This spin cutoff becomes a phenomenological parameter related to the inflation energy scale detectable via CMB observations.\footnote{The author is grateful to R.Metsaev for the discussion of this point.}

Thus, the CHS models dominated by fermions with $B<0$ and anomaly free models with $B=0$ remain a playground of holographic methods in AdS/CFT correspondence, but fail to generate interesting CFT driven cosmology. This matches with the fact established in \cite{DGP/CFT} that the CFT driven scenario implies another type of holographic duality -- the DGP/CFT correspondence. For positive $\beta\sim B>0$ this is the duality between the 4-dimensional CFT driven cosmology of \cite{slih} and the tree-level dynamics of the 4D brane which is embedded into the 5-dimensional Schwarzschild-de Sitter bulk and carries the 4D Einstein-Hilbert term -- the generalization of the DGP model \cite{DGP}.

\section*{Acknowledgements}
I would like to express my gratitude to A.Tseytlin, M.Vasiliev and M.Beccaria for fruitful and thought-provoking correspondence and discussions. I also greatly benefitted from helpful discussions with F.Bezrukov, D.Blas, C.Burgess, C. Deffayet, J.Garriga, J.B.Hartle, A.Yu.Kamenshchik, D.Marolf, R.Metsaev, D.V.Nesterov, M.Shaposhnikov, S.Sibiryakov, N.Tsamis, W.Unruh, A.Vilenkin and R.Woodard. I am grateful for hospitality of Theory Division, CERN, and Pacific Institute of Theoretical Physics, UBC, where this work was initiated. This work was also supported by the RFBR grant No. 14-01-00489 and by the Tomsk State University Competitiveness Improvement Program.

\end{document}